\documentclass[epsf,showpacs,twocolumn,prd,aps]{revtex4}
\usepackage{amssymb}
\usepackage{amsmath}

\usepackage[dvips]{epsfig}

\begin{document}

\title{Graviton mediated photon-photon scattering in general relativity}
\author{Gert Brodin,\footnote{Also at: Centre for Fundamental Physics, Rutherford Appleton Laboratory, Chilton, Didcot, Oxon OX11 OQX, U.K.} Daniel Eriksson, and Mattias Marklund$^*$}
\affiliation{Department of Physics, Ume{\aa} University SE--901 87 Ume{\aa}, Sweden}
\date{\today}

\begin{abstract}
In this paper we consider photon-photon scattering due to self-induced
gravitational perturbations on a Minkowski background. We focus on four-wave
interaction between plane waves with weakly space and time dependent
amplitudes, since interaction involving a fewer number of waves is excluded
by energy-momentum conservation. The Einstein-Maxwell system is solved
perturbatively to third order in the field amplitudes and the coupling
coefficients are found for arbitrary polarizations in the center of mass
system. Comparisons with calculations based on quantum field theoretical
methods are made, and the small discrepances are explained.
\end{abstract}

\maketitle


\section{Introduction}

As is wellknown, photon-photon scattering can occur due to the exchange of
virtual electron-positron pairs, as described by QED or modifications
thereof, see e.g.\ \cite
{QED-general}, and may even lead to collective photon phenomena \cite{collective}. 
Photon interactions via the quantum vacuum, sometimes involving deviations from the standard model,
has recently been much discussed in the literature due to advances in 
experimental technologies (see e.g.\ \cite{exp}) as well as new theoretical insights \cite{new}.  
Moreover, photons also interact gravitationally, although
this effect has been much less studied. Purely general relativistic
treatments of electromagnetic wave interactions have been made resulting in
exact solutions, see e.g. \cite{Griffiths91}, but these calculations are
very different from the pure scattering processes, and do not address the
interaction at the single photon level. On the other hand, it is not clear
to what extent calculations of the gravitational cross-section using quantum
field theoretical methods \cite{Barker1966,Barker1967} (see also \cite{Dewitt}) are consistent with
classical general relativity. In order to shed light on this issue, we will
consider the interaction of four electromagnetic (EM) waves on a Minkowski
background, which is the lowest order scattering process consistent with
energy-momentum conservation. By studying the classical Einstein-Maxwell
system, but ignoring terms that do not correspond to pure scattering (e.g.
frequency shift terms) we will attempt to make contact between the classical
and quantum field theoretical picture. Calculating the classical coupling
coefficients between waves of different polarizations, corresponding to the
scattering amplitudes in quantum field theory, we are able to compare the
classical cross-section with that of quantum field theory \cite
{Barker1966,Barker1967}. While the results are approximately equal for small
scattering angles $\theta $, we find that there are significant differences
for large $\theta $. The likely source behind this discrepancy is that the
quantum field theoretical calculation \cite{Barker1967} used the matrix
scattering amplitude in order to define the interaction potential. As shown
by Ref. \cite{Kazakov2001}, such a procedure is not able to fully reproduce
the general relativistic potential.

Finally we note that while gravitational photon-photon scattering is weaker
than the QED scattering in most cases of physical interest \cite{QED-general,collective}%
, it should be noted that in the long wavelength limit, actually
the gravitational cross-section is larger than that due to QED. 

\section{Theory and results}

We employ units such that the speed of light and Planck's constant are $c= \hbar =1$, and use metric
signature $\left( -,+,+,+\right) $. Tetrad indices $a,b,\ldots $ run from 0
to 3 and $\alpha ,\beta ,\ldots $ from 1 to 3. Coordinate indices $\mu ,\nu
,\ldots $ go from 0 to 3.\newline
Assuming plane waves and denoting the interacting waves $A$, $B$, $C$ and $D$
the total electric field is given by 
\[
E=\sum_{n}\left[ E_{n}(x^{\mu })e^{ik_{n\mu }x^{\mu }} + \mathrm{c.c.}\right] \hspace{%
2mm},\hspace{2mm}n=A,B,C,D ,
\]
where c.c.\ denotes the complex conjugate. A similar expression
holds for the magnetic field. Moreover we assume that the amplitudes have a
weak dependence on space and time, i.e. $\left| \partial _{\mu }E\left(
x^{\nu }\right) \right| \ll \left| k_{\mu }\right| \left| E\left( x^{\nu
}\right) \right| $. The presence of the EM fields induce a perturbation, $%
h_{\mu \nu }$, of the flat background metric, $\eta _{\mu \nu }$, enabling
energy exchange between the modes. A generic frame, orthonormal to quadratic
order in the field amplitudes (linear order in the metric perturbation), is
chosen as 
\begin{eqnarray}
e_{0}^{\mu } &=&\left( 1+\frac{1}{2}h_{00},-\frac{1}{2}h_{01},-\frac{1}{2}%
h_{02},-\frac{1}{2}h_{03}\right)  , \nonumber \\
e_{1}^{\mu } &=&\left( \frac{1}{2}h_{01},1-\frac{1}{2}h_{11},-\frac{1}{2}%
h_{12},-\frac{1}{2}h_{13}\right) , \nonumber \\
e_{2}^{\mu } &=&\left( \frac{1}{2}h_{02},-\frac{1}{2}h_{12},1-\frac{1}{2}%
h_{22},-\frac{1}{2}h_{23}\right)  , \nonumber \\
e_{3}^{\mu } &=&\left( \frac{1}{2}h_{03},-\frac{1}{2}h_{13},-\frac{1}{2}%
h_{23},1-\frac{1}{2}h_{33}\right) \hspace{2mm},
\end{eqnarray}
where $\mathbf{e}_{a}=e_{a}^{\mu }\partial _{\mu }$. The matching condition
corresponding to energy and momentum conservation is given by 
\begin{equation}
k_{A}^{\mu }+k_{B}^{\mu }=k_{C}^{\mu }+k_{D}^{\mu }\hspace{2mm}.
\label{matching}
\end{equation}
In the center of mass system all frequencies are equal and the waves are
counterpropagating pairwise, i.e. the wave vectors satisfy $\mathbf{k}_{A}=-%
\mathbf{k}_{B}$, $\mathbf{k}_{C}=-\mathbf{k}_{D}$. We make the following
ansatz for the metric perturbations 
\begin{eqnarray*}
&&\!\!\!\!\!\!\! h_{\alpha \beta } = \bar{h}_{\alpha \beta}\left( x,z,t\right) e^{-2i\omega
t} + \bar{h}_{\alpha \beta}^{\ast }\left( x,z,t\right) e^{2i\omega t} 
\\
&& \!\!\!\!\!\!  + \tilde{h}_{\alpha \beta}\left( t,x,z\right) e^{-i\left( \mathbf{k}_{A}+%
\mathbf{k}_{C}\right)\cdot \mathbf{x}} + \tilde{h}_{\alpha \beta}^{\ast }\left(
t,x,z\right) e^{i\left( \mathbf{k}_{A}+\mathbf{k}_{C}\right)\cdot \mathbf{x}} 
 \\
&& \!\!\!\!\!\!  + \hat{h}_{\alpha \beta}\left( t,x,z\right) e^{-i\left( \mathbf{k}_{A}-%
\mathbf{k}_{C}\right)\cdot \mathbf{x}} + \hat{h}_{\alpha \beta}^{\ast }\left(
t,x,z\right) e^{i\left( \mathbf{k}_{A}-\mathbf{k}_{C}\right)\cdot \mathbf{x}}%
. 
\end{eqnarray*}
This is not the most general ansatz, however it is sufficient to give all
terms corresponding to resonant energy exchange between the modes. All 
the components $\bar{h}_{\alpha \beta}$, $\bar{h}_{\alpha \beta}^{\ast }$, $\tilde{h}_{\alpha \beta
}$, $\tilde{h}_{\alpha \beta }^{\ast }$, $\hat{h}_{\alpha \beta}$, $%
\hat{h}_{\alpha \beta}^{\ast }$, except twelve of them, are determined by the field
equations, $G_{ab}=\kappa T_{ab}$, where $G_{ab}$ is the Einstein tensor and 
$T_{ab}$ the energy-momentum tensor. In the present case the only
contribution to the energy-momentum tensor is given by 
\[
T_{ab}=F_{a}^{c}F_{bc}-\tfrac{1}{4}g_{ab}F^{cd}F_{cd}\hspace{2mm}. 
\]
The undetermined coefficients in the metric ansatz are set to zero using the
generalized Lorentz condition. Note that the non-zero coefficients are of
quadratic order in the field amplitudes. Using Maxwell's equations 
$\nabla _{[a}F_{bc]} = 0$ and $\nabla _{a}F^{ab} =j^{b}$,
where $F^{ab}$ is the electromagnetic field tensor and $j^{b}$ the
four-current density, we can derive the following wave equations 
\begin{eqnarray}
\tilde{\square}E^{\alpha } &=&-\mathbf{e}_{0}j_{E}^{\alpha }-\epsilon
^{\alpha \beta \gamma }\mathbf{e}_{\beta }j_{B\gamma }-\delta ^{\alpha
\gamma }\mathbf{e}_{\gamma }\rho _{E}-  \nonumber \\
&&\epsilon ^{\alpha \beta \gamma }C_{\beta 0}^{a}\mathbf{e}_{a}B_{\gamma
}-\delta ^{\alpha \gamma }C_{\beta \gamma }^{a}\mathbf{e}_{a}E^{\beta }%
\hspace{6mm}  , \label{we1} \\
\tilde{\square}B^{\alpha } &=&-\mathbf{e}_{0}j_{B}^{\alpha }+\epsilon
^{\alpha \beta \gamma }\mathbf{e}_{\beta }j_{E\gamma }-\delta ^{\alpha
\gamma }\mathbf{e}_{\gamma }\rho _{B}+  \nonumber \\
&&\epsilon ^{\alpha \beta \gamma }C_{\beta 0}^{a}\mathbf{e}_{a}E_{\gamma
}-\delta ^{\alpha \gamma }C_{\beta \gamma }^{a}\mathbf{e}_{a}B^{\beta }.%
\hspace{5mm}
\end{eqnarray}
Here the wave operator $\tilde{\square}\equiv \mathbf{e}_{0}\cdot\mathbf{e}_{0} 
+ \nabla \cdot \nabla $, which coincides with the D'Alembertian operator
in Euclidian space, and $C_{ab}^{c}$ are commutation functions for the frame
vectors satisfying $\left[ \mathbf{e}_{a},\mathbf{e}_{b}\right] =C_{ab}^{c}%
\mathbf{e}_{c}$. $\mathbf{j}_{E}$, $\mathbf{j}_{B}$, $\rho _{E}$ and $\rho
_{B}$ are the effective currents and charges due to the inclusion of the
gravitational field given by 
\begin{eqnarray}
\mathbf{j}_{E} &=&\Big[ -\left( \Gamma _{0\beta }^{\alpha }-\Gamma _{\beta
0}^{\alpha }\right) E^{\beta }+\Gamma _{0\beta }^{\beta }E^{\alpha }- 
\nonumber \\
&& \epsilon ^{\alpha \beta \gamma }\left( \Gamma _{\beta
0}^{0}B_{\gamma }+\Gamma _{\beta \gamma }^{\delta }B_{\delta }\right) \Big]
\mathbf{e}_{\alpha }\hspace{2mm}, \\
\mathbf{j}_{B} &=&\Big[ -\left( \Gamma _{0\beta }^{\alpha }-\Gamma _{\beta
0}^{\alpha }\right) B^{\beta }+\Gamma _{0\beta }^{\beta }B^{\alpha }- 
\nonumber \\
&& \epsilon ^{\alpha \beta \gamma }\left( \Gamma _{\beta
0}^{0}E_{\gamma }+\Gamma _{\beta \gamma }^{\delta }E_{\delta }\right) \Big]
\mathbf{e}_{\alpha }\hspace{2mm}, \\
\rho _{E} &=&-\Gamma _{\beta \alpha }^{\alpha }E^{\beta }-\epsilon ^{\alpha
\beta \gamma }\Gamma _{\alpha \beta }^{0}B_{\gamma }\hspace{2mm}, \\
\rho _{B} &=&-\Gamma _{\beta \alpha }^{\alpha }B^{\beta }-\epsilon ^{\alpha
\beta \gamma }\Gamma _{\alpha \beta }^{0}E_{\gamma }\hspace{2mm},
\end{eqnarray}
where $\Gamma _{bc}^{a}$ are the Ricci rotation coefficients. The effective
currents and charges will be of cubic order in the field amplitudes.
Eliminating the magnetic field from the wave equation (\ref{we1}) by using
Faraday's law to leading order, $\mathbf{k}\times \mathbf{E}=\omega \mathbf{B%
}$, and neglecting terms of order four or higher in the field amplitudes
will result in three categories of terms.

\begin{enumerate}
\item  {Nonresonant terms that will vanish after averaging over several
wavelengths and time periods. Interaction due to these terms are not
consistent with energy momentum conservation. Most terms belong in this
category.}

\item  {Phase shift terms which are resonant but give rise to phase shifts
rather than scattering. These terms typically contain a certain amplitude
together with its complex conjugate, i.e. terms of the form $%
E_{A}E_{A}^{\ast }E_{B}$.}

\item  {Resonant scattering terms containing all three wave amplitudes
according to the energy-momentum conservation condition (\ref{matching}).}
\end{enumerate}

Based on the classification of terms above we restrict ourselves to include
only the resonant scattering terms and introduce polarization states
perpendicular to the wave vectors as shown in figure 1. Note that the $E_{2}$
and the $E_{\times }$ directions coincide. We thus have 
\begin{eqnarray}
&& 
E_{A1} =-\cos \left( {\theta }/{2}\right) E_{A+}, \quad 
E_{A3}=\sin \left( {\theta }/{2}\right) E_{A+}  , \nonumber \\
&& 
E_{B1} =\cos \left( {\theta }/{2}\right) E_{B+}, \quad \,\,\,\,\,
E_{B3}=-\sin \left( {\theta }/{2}\right) E_{B+}  ,  \nonumber \\
&&
E_{C1} =-\cos \left( {\theta }/{2}\right) E_{C+}, \quad 
E_{C3}=-\sin \left( {\theta }/{2}\right) E_{C+}  ,
\nonumber \\
&& E_{D1} =\cos \left( {\theta }/{2}\right) E_{D+}, \quad  \,\,\,\,\,
E_{D3}=\sin \left( {\theta }/{2}\right) E_{D+} ,  \nonumber
\end{eqnarray}
with $E_{A2}=E_{A\times}, E_{B2}=E_{B\times}, E_{C2}=E_{C\times }, E_{D2}=E_{D\times}$,
and
\begin{eqnarray}
k_{Ax}&=&\sin\left({\theta}/{2}\right)\omega ,\hspace{5mm}%
k_{Az}=\cos\left({\theta}/{2}\right)\omega  , \nonumber \\
k_{Bx}&=&-\sin\left({\theta}/{2}\right)\omega ,\hspace{2mm}%
k_{Bz}=-\cos\left({\theta}/{2}\right)\omega ,  \nonumber \\
k_{Cx}&=&-\sin\left({\theta}/{2}\right)\omega ,\hspace{2mm}%
k_{Cz}=\cos\left({\theta}/{2}\right)\omega  , \nonumber \\
k_{Dx}&=&\sin\left({\theta}/{2}\right)\omega ,\hspace{5mm}%
k_{Dz}=-\cos\left({\theta}/{2}\right)\omega . \nonumber
\end{eqnarray}

\begin{figure}[tbp]
\includegraphics[width=0.75\columnwidth]{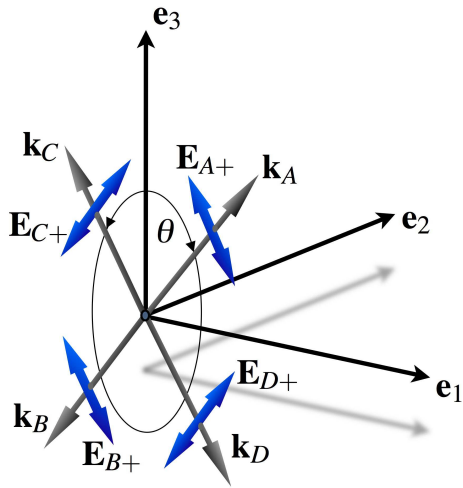}
\caption{Polarization directions and the definition of the scattering angle $\theta$.}
\label{polfig}
\end{figure}
After some lengthy but straightforward algebra we end up with the following
coupling equations describing the evolution of the wave amplitudes to
leading order 
\begin{eqnarray}
\square E_{A+} &=&F_{1}E_{B+}^{\ast }E_{C+}E_{D+}+F_{2}E_{B+}^{\ast
}E_{C\times }E_{D\times }  \nonumber \\
&&+F_{3}E_{B\times }^{\ast }E_{C\times }E_{D+}+F_{4}E_{B\times }^{\ast
}E_{C+}E_{D\times }  , \label{deap}
\end{eqnarray}
\begin{eqnarray}
\square E_{B+} &=&F_{1}E_{A+}^{\ast }E_{C+}E_{D+}+F_{2}E_{A+}^{\ast
}E_{C\times }E_{D\times }  \nonumber \\
&&+F_{3}E_{A\times }^{\ast }E_{C+}E_{D\times }+F_{4}E_{A\times }^{\ast
}E_{C\times }E_{D+} , \label{debp}
\end{eqnarray}
\begin{eqnarray}
\square E_{C+} &=&F_{1}E_{D+}^{\ast }E_{A+}E_{B+}+F_{2}E_{D+}^{\ast
}E_{A\times }E_{B\times }  \nonumber \\
&&+F_{3}E_{D\times }^{\ast }E_{A\times }E_{B+}+F_{4}E_{D\times }^{\ast
}E_{A+}E_{B\times }  , \label{decp}
\end{eqnarray}
\begin{eqnarray}
\square E_{D+} &=&F_{1}E_{C+}^{\ast }E_{A+}E_{B+}+F_{2}E_{C+}^{\ast
}E_{A\times }E_{B\times }  \nonumber \\
&&+F_{3}E_{C\times }^{\ast }E_{A+}E_{B\times }+F_{4}E_{C\times }^{\ast
}E_{A\times }E_{B+} ,\label{dedp}
\end{eqnarray}
where $\square =\partial ^{2}/\partial t^{2}-\partial ^{2}/\partial
x^{2}-\partial ^{2}/\partial z^{2}$ and 
\begin{eqnarray}
&& F_{1} =\frac{\kappa \left( 3+\cos ^{2}\theta\right)^2 }{%
1-\cos ^{2}\theta },\hspace{2mm} \nonumber \\
&& F_{2}=-\kappa \left( 7+\cos ^{2}\theta
\right) ,  \nonumber \\
&&F_{3} =\frac{4\kappa \left( 2+\cos ^{2}\theta +\cos \theta \right) }{%
1+\cos \theta },\hspace{2mm} \nonumber \\ 
&& F_{4}=\frac{4\kappa \left( 2+\cos ^{2}\theta
-\cos \theta \right) }{1-\cos \theta } .  \label{E-coupling}
\end{eqnarray}
For symmetry reasons $\square E_{A\times },\square E_{B\times },\square
E_{C\times }$ and $\square E_{D\times }$ can be found from (\ref{deap})--(\ref
{dedp}) respectively by interchanging $+$ and $\times $. The coupling
coefficients only depend on the scattering angle $\theta $, and in the limit
when $\theta \rightarrow 0$ both $F_{1}$ and $F_{4}$ become infinite while $%
F_{2}\rightarrow -F_{3}$. The small angle divergence in $F_{1}$ and $F_{4}$ is a 
consquence of the infinite range of the gravitational force. However, the coefficients 
$F_{2}$ and $F_{3}$ must remain finite for all angles, as those coefficients not only 
describe scattering an angle $\theta$, but also correspond to a change in the polarization state. 

In order to check the consistency of our results, we assume long pulses, i.e.\ $\square \approx -2i\omega \partial _{t}$,
such that the time derivative of the total energy density, $\varepsilon _{tot}=\sum_n
( | E_{n+}|^2 + | E_{n\times }|^2 ) $ (where the sum is over $A, B, C, D$), can
be easily calculated. Carrying out the sum, it is found that all the scattering terms cancel, and 
thus we deduce that the evolution equations (\ref{deap})-(\ref
{dedp}) are energy conserving.

Next we rewrite (\ref{deap})--(\ref{dedp}) in terms of the vector potentials,
which rescales the coupling coefficients (\ref{E-coupling}) by a factor $%
\omega ^{2}$. Noting that the rescaled coupling coefficients corresponds to
the scattering amplitudes, and following Ref. \cite{Itzykson85}, we find
that the unpolarized differential cross-section can be calculated as 
\begin{equation}
\frac{\partial \sigma }{\partial \Omega }=\frac{\left| \mathcal{M}\right|
^{2}}{128\omega ^{2}\left( 2\pi \right) ^{2}}\hspace{2mm},
\label{Cross-section}
\end{equation}
where the square of the scattering matrix amplitude averaged over all
polarization states is given by

\begin{eqnarray}
\left| \mathcal{M}\right| ^{2} &=&\frac{\omega ^{4}\kappa ^{2}}{\sin
^{4}\theta }\left( \cos ^{8}\theta +28\cos ^{6}\theta +70\cos ^{4}\theta
\right.   \nonumber \\
&&\qquad\qquad \left. + 28\cos ^{2}\theta +129\right) .   \label{differential-expression}
\end{eqnarray}
This result should be compared with the same quantity calculated by quantum
field theoretical methods, i.e. Eq. (15) in Ref. \cite{Barker1967}. It turns
out that the differential cross-sections agree in the limit $\theta \ll \pi
/2$, but as seen from Fig 2, where the classical and quantum field
theoretical expressions are shown, the two expressions differ slightly in
general. In order to resolve the difference more accurately, the ratio of the
cross-sections are shown in Fig. 3, where one should note the agreement for small angles.  
However, for general angles the expressions clearly disagree, and it is natural to ask what causes
this discrepancy. To answer this question we note that Ref. \cite{Barker1967}
has used the matrix scattering amplitude to determine the interaction
potential. As demonstrated by Ref. \cite{Kazakov2001}, however, such a
procedure is not sufficient to fully reproduce the general relativistic
potential. As the general relativistic deviation from Newtonian behavior
becomes more pronounced for large scattering angles, this explains the
deviation for general angles, but also the agreement in the small angle
limit. 
\begin{figure}[tbp]
\centering
\includegraphics[width=0.9\columnwidth]{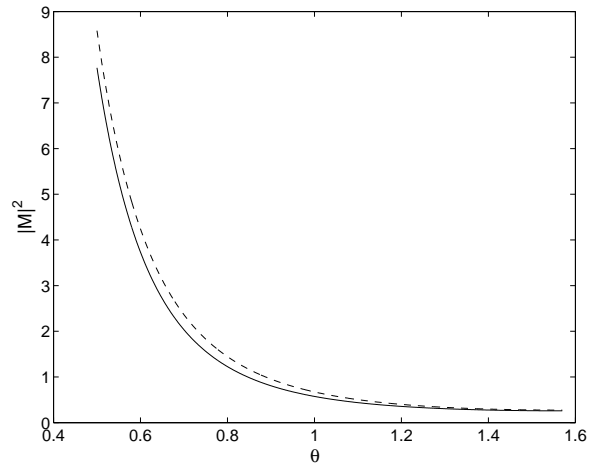}
\caption{The classical(solid line)and quantum field theoretical (dashed
line) cross-sections in arbitrary units as a function of scattering angle.}
\end{figure}

\begin{figure}[tbp]
\centering
\includegraphics[width=0.9\columnwidth]{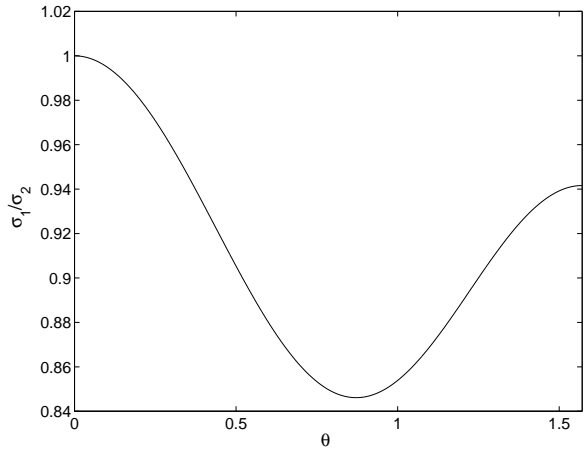}
\caption{The ratio of the classical and quantum field theoretical
cross-sections $\sigma_1$ and $\sigma_2$, respectively, as a function of scattering angle.}
\end{figure}

\section{Summary and Discussion}

Comparing the calculated cross-section for gravitational photon-photon
scattering Eq. (\ref{Cross-section}) with that from QED photon-photon
scattering (due to exchange of virtual electron-positron pairs), we find
that they have different frequency dependence. The former is proportional to 
$\omega ^{2}$ while the latter is proportional to $\omega ^{6}$ \cite
{Itzykson85}. 
Noting from Eq. (\ref{Cross-section}) that $\partial \sigma /\partial \Omega\approx 16 (2\pi)^2 L_p^4/\lambda^2$
(letting $\sin\theta\sim\cos\theta\sim1/\sqrt{2}$), where $L_p$ is the Planck length  and $\lambda$ the wavelength, 
and comparing with the QED expression for the cross-section (e.g. Ref. \cite{Itzykson85}), we find that 
the QED and gravitational cross-sections are comparable for frequencies
\begin{equation}
\omega \sim10^{3}c\left({L_p^2}/{r_0\lambda_c^3}\right)^{1/2} ,
\end{equation}
where $r_0$ is the classical electron radius, $\lambda_c$ is the Compton wavelength,
and we have reinstated the speed of light $c$. Thus 
the gravitational effects become the dominant contribution to the cross-section for
frequencies $\omega\sim 30\, \rm{rad/s}$
and lower. Still, the cross-section is
very small, and we need extremely large photon densities for gravitational
photon-photon scattering to influence the dynamics. Situations that could be
of interest to study in more detail involve the dense photon gas
surrounding pulsars \cite{Kondratyev}, as well as the photon
gas in the early universe \cite{Kolb-Turner}.
Furthermore, we note that if the energy densities are sufficiently high \cite
{Sufficient-note}, the timescales for nonlinear evolution will not be
determined by the cross-section, even if the spectrum is strongly
incoherent. Instead the characteristic time-scale must be found from weak
turbulence theories \cite{Hasegawa-book}. Using the so called random phase
approximation, the phase dependence can be integrated out, and evolution
equations for the spectral energy densities are derived \cite{Hasegawa-book}.

Graviton mediated photon-photon scattering share many parameter
similarities with photon-graviton pair conversion \cite{Andreas}. 
While it is possible that gravitational photon-photon scattering may have
applications to astrophysics and/or cosmology, the effect can typically be
neglected compared to other effects, such as QED photon-photon scattering
or, in the presence of matter, interaction with charged particles. Thus our
main aim here has been to make an explicit comparison with the quantum
field theoretical result (see Fig. 2), which show a slight deviation from
our general relativistic cross-section. As seen in Fig. 3, the deviation
vanishes in the limit of small scattering angles. We trace the difference
between the quantum field theoretical and the classical result to the
difficulty in determining the general relativistic interaction potential
from the matrix scattering amplitude, as done by Ref. \cite{Barker1967}. An
interesting problem, which is a project for further research, is to
investigate whether a quantum field theoretical calculation can be improved
to incorporate a fully general relativistic interaction potential.


\begin{thebibliography}{9}
\bibitem{Sufficient-note}  A ''sufficiently high'' photon density here means
that the number of photons present in a cube of size $\lambda _{\mathrm{ch}%
}^{3}$, where $\lambda _{\mathrm{ch}}$ is the characteristic wavelength of
the spectrum, is much larger than unity. Note that for a photon gas subject
to cosmological \ expansion, this number is a constant. Furthermore, for the
particular case of the cosmic background radiation, this constant is much
smaller than unity.
\end{thebibliography}
\end{document}